\documentclass[12pt]{article}
\usepackage[pdftex]{graphicx}

\begin{document}
\date{}
\title{Noise spectroscopy of optical microcavity}
\maketitle

 \hskip60mm G. G. Kozlov \vskip3mm

\hskip100pt {\it e}-mail:  gkozlov@photonics.phys.spbu.ru

Institute of Physics, Saint-Petersburg State University, Spin
Optics Laboratory, Saint-Petersburg, 198504, Russia \vskip10mm

\begin{abstract}
The intensity noise spectrum of the light passed through an
optical microcavity is calculated with allowance for thermal
fluctuations of its thickness. The spectrum thus obtained reveals
a peak at the frequency of acoustic mode localized inside the
microcavity and depends on the size of the illuminated area. The
estimates of the noise magnitude show that it can be detected
using the up-to-date noise spectroscopy technique.
\end{abstract}

\section{Introduction}

The recent progress in the area of digital high-speed spectrum analyzers of electric signals gave rise
to a strong increase of the interest to the optical noise spectroscopy
\cite{Yab,Mitsui,Crooker,McIntyre,Walser,Oestreich}. In the first experimental
observation of EPR of sodium atoms in the Faraday rotation noise spectrum \cite{Zap}, a high polarimetric
sensitivity \cite{Zap1} made it possible to detect the above spectrum using a conventional lock-in amplifier.
Application of the up-to-date digital Fourier analyzers with accumulating systems of data acquisition
allowed one to detect magnetic resonance spectra in a number
 of solid-state objects \cite{Crooker} and to widen
the spectral range of the noise signal detection up to several GHz. In this note, we describe the
effect of the noise modulation of intensity of quasi-monochromatic light passed through an optical
microcavity. The modulation arises due to thermal fluctuations of thickness (and, hence, of the
resonance frequency) of the microcavity. In accordance with the calculations presented in this work,
experimental observation of this effect proves to be possible with the aid of the digital spectrum
analyzer mentioned above. In this paper, we primarily consider possibility of detection of this effect
and only briefly discuss its informative potentialities.

Let us clarify in more detail the idea of the proposed effect. Consider an optical cavity (Fabry-Perot
interferometer) - two mirrors separated by a gap $L$. When the reflectivity of the mirrors is close to
unity, the frequency dependence of the transmission coefficient $I/I_0$ of such a cavity
 in the vicinity of
the principal mode can be written in the form
\begin{equation}
{I\over I_0}={\Delta^2\over \Delta^2+(\omega-\omega_0)^2}
\label{1}
\end{equation}
Here, $I_0 (I)$ is the intensity of the monochromatic plane wave with the frequency $\omega$ at the
entrance (at the exit) of the interferometer and $\omega_0=\pi c/L$
 is the resonance frequency of the interferometer
($c$ is the speed of light in the medium between the mirrors). The width of the transmission spectrum
of the interferometer $\Delta$ is determined by the reflectivities of the mirrors and, for real
microcavities, the finesse $Q\equiv\omega_0/\Delta$ can reach 1000.
 A change in the cavity thickness $L\rightarrow L+\xi$ gives rise to a change
in its resonance frequency   $\omega_0\rightarrow\omega_0+\delta\omega_0$,
which, at  $\xi/L<<1$, is described by the relationship
 $|\delta\omega_0/\omega_0|=|\xi/L|$. Bearing this in
mind, one can easily make sure, using (\ref{1}), that the relative change of the cavity thickness
 $|\xi/L|\sim 1/Q$,
results in changes of its transmission coefficient of about unity and can be easily detected. In this
case, the absolute changes of the cavity thickness $\xi$, for typical values of parameters of the optical
microcavities, $L = 0.25 \mu$m, $Q$ = 1000, are of the order of atomic dimensions $\xi\sim 0.25$ nm.
 So high sensitivity of
transmission of the optical cavity to variations of its parameters is used in physical experiments
\cite{Hu}
and allows one to set a question about possibility to detect  the cavity
transmission  noise related to the cavity thickness thermal fluctuations. To evaluate possibility of
observation of this noise, its magnitude should be compared with that of the shot noise of the used
light source. Corresponding calculations are presented in the next section.

\section{Model calculations}

Consider an optical cavity formed by a thin layer of a medium, with the thickness $L$, sandwiched
 between  two
thin reflecting films. Let a monochromatic light beam with the intensity $I_0$ and frequency $\omega$ be
incident on this cavity. Denote the area of the light spot on the cavity as $D^2$ and the cavity
finesse as $Q$. \footnote{The finesse measured by the half-width of the transmission spectrum may
depend on the transverse beam size $D$, but, at normal incidence, this dependence is weak} The
coordinate system is chosen so that the plane of the cavity coincides with the $xy$ plane. Then, the
cavity thickness will be a function of $x$ and $y$, which can be presented as a sum of the constant mean
thickness $L$ and small thermal fluctuation $\xi(t,x,y)$. We assume that the resonance frequency of the
cavity is determined by its thickness averaged over the area $D^2$ of the light spot. \footnote{In
other words, this is the frequency at which the transmission coefficient of the cavity is the
greatest.} Now, the fluctuation $\delta I$ of the transmitted light intensity can be written as
   \begin{equation}
   \delta I=G(\omega)\int_{D^2} dx dy\hskip1mm\xi(t,x,y),\hskip2mm\hbox{ where }\hskip1mm
   G(\omega)={I_0\over D^2}{d\over d\omega_0}\bigg({{\Delta^2\over \Delta^2+(\omega-\omega_0)^2}}\bigg)\hskip1mm
   {d\omega_0\over d\xi}\bigg|_{\xi=0},\hskip2mm \omega_0={\pi c\over L+\xi}.
   \label{3}
   \end{equation}
The greatest value of the factor $G$ is attained at  $\omega=\omega_0-\Delta$:
$G(\omega_0+\Delta)={I_0Q/ 2LD^2}$. We will be interested in the intensity noise
spectrum $S(\nu)$ of the light transmitted by the cavity. The function $S(\nu)$
 is connected with the correlation
function  $\langle \delta I(0)\delta I(t)\rangle$ by the  relationship:
\begin{equation}
 S(\nu)=\int dt e^{\imath\nu t}\langle \delta I(0)\delta I(t)\rangle
\label{4}
\end{equation}
Using (\ref{3}), we obtain for $\langle \delta I(0)\delta I(t)\rangle$ the following expression:
\begin{equation}
\langle \delta I(0)\delta I(t)\rangle
=G^2(\omega)\int_{D^2}\int_{D^2} dxdydx'dy'\langle\xi(0,x,y)\xi(t,x',y')\rangle
\label{5}
\end{equation}

To calculate the correlation function $\langle\xi(0,x,y)\xi(t,x',y')\rangle$ entering Eq.(\ref{5}) we have
to:

(i)     specify a model of motion of the material of the cavity,

(ii)    obtain the appropriate Hamiltonian, and

(iii)   fulfill averaging in Eq.(\ref{5}) with the thermodynamically equilibrium distribution function
$\sim\exp[-H/kT]$.

We will describe dynamics of the cavity material in terms of small acoustic waves.
We assume that the cavity occupies the region
$z\in[0,L]$,  $x\in[0,a]$, and $y\in[0,b]$.
Since the optical transmission spectrum depends on the cavity thickness along the $z$-axis, we will be
interested only in $z$-projection of displacement of the cavity material. According to the adopted
model, this displacement is described by the acoustic field $u(x,y,z)$ satisfying the wave equation
 \begin{equation}
 {\partial^2 u\over \partial t^2}=v^2\Delta u,
\label{6}
 \end{equation}
Where $v$ -- is the velocity of longitudinal sound in the cavity material. The energy of the
acoustic  field  $u(x,y,z)$ is given by the formula
 \begin{equation}
 E=\int dxdydz\bigg\{{\rho \dot u^2\over 2}+{\gamma\over 2}\bigg[\bigg({\partial u\over \partial x}\bigg)^2+
\bigg({\partial u\over \partial y}\bigg)^2+\bigg({\partial u\over \partial z}\bigg)^2\bigg]
 \bigg\}
 \label{8}
 \end{equation}
Here, $\rho$ -- is the density of the cavity material and $\gamma$
 --  is the constant that described the elastic strain
energy density. Connection of this constant with the velocity of sound $v$ will be given below. The
direct substitution shows that expansion of solution of Eq. (\ref{6}) in terms of normal modes, satisfying
the boundary conditions
\begin{equation}
{\partial u\over \partial x}\bigg|_{x=0,a}=0,\hskip1mm
{\partial u\over \partial y}\bigg|_{y=0,b}=0,\hskip1mm
{\partial u\over \partial z}\bigg|_{z=0,L}=0,
\label{7}
\end{equation}
which correspond to zero strains at the bounds of the cavity, has the form:
  \begin{equation}
  u(t,x,y,z)=\sum_{pmn} u_{pmn}\cos\bigg({\pi p x\over a}\bigg)\hskip1mm
\cos\bigg({\pi m y\over b}\bigg)\hskip1mm\cos\bigg({\pi n z\over L}\bigg),\hskip2mm
p,n,m \hbox{ integers }
\hskip2mm p,m,n>0
\label{9}
  \end{equation}
where {\it the degrees of freedom} of the acoustic field $u_{pmn}$ meet the following equations of motion
\begin{equation}
 \ddot u_{pmn}=-\omega_{pmn}^2 u_{pmn},\hskip2mm\hbox{ where }\hskip2mm
 \omega_{pmn}^2=v^2\bigg[\bigg({\pi p \over a}\bigg)^2+\bigg({\pi m \over b}\bigg)^2+
 \bigg({\pi n \over L}\bigg)^2\bigg]
 \label{10}
 \end{equation}
Using Eq.(\ref{9}), we can express the energy $E$ (\ref{8}) through the degrees of freedom
 $u_{pmn}$:
\begin{equation}
E={V\over 16}\sum_{pmn}\bigg[
\rho \dot u_{pmn}^2+\gamma u_{pmn}^2\bigg[
\bigg({\pi p\over a}\bigg)^2+\bigg({\pi m\over b}\bigg)^2+\bigg({\pi n\over L}\bigg)^2\bigg]
\bigg]=
\label{12}
\end{equation}
$$
= {V\over 16}\sum_{pmn}\bigg[
\rho \dot u_{pmn}^2+{\gamma\omega_{pmn}^2\over v^2} u_{pmn}^2\bigg]
$$
To obtain the Hamiltonian corresponding to energy (\ref{12}) , we have to introduce the generalized momenta
$p_{pmn}$, conjugated with the generalized coordinates $u_{pmn}$ so that the
equations of motion (\ref{10}) acquire the form of the Hamilton equations:
  \begin{equation}
  {\partial H\over \partial u_{pmn}}=-\dot p_{pmn}
\hskip10mm
  {\partial H\over \partial p_{pmn}}=\dot u_{pmn}
\label{13}
  \end{equation}
If we set
$$
p_{pmn}\equiv {\dot u_{pmn}\rho V\over 8}\hskip4mm\hbox{ and }
 \hskip4mm  \gamma=\rho v^2,
$$
and express energy (\ref{12}) through  $p_{pmn}$ and $u_{pmn}$:
\begin{equation}
H=\sum_{pmn}\bigg[{4p_{pmn}^2\over M}+{M\omega_{pmn}^2 u_{pmn}^2\over 16}\bigg]
\hskip3mm\hbox{ где }\hskip3mm M\equiv \rho V\hskip3mm\hbox{--  resonator mass,}
\hskip4mm p_{pmn}={M\over 8}\dot u_{pmn},
\label{15}
\end{equation}

then we can easily make sure that equations of motion (\ref{10}) are equivalent to Eq. (\ref{13}).
Thus, Eq.(\ref{15}) is the sought Hamiltonian.

Using Hamiltonian (\ref{15}), we can write the following expression for the distribution function
  $\sigma\{p_{pmn},u_{pmn}\}$
 of the generalized coordinates and momenta in the state of thermodynamic equilibrium with the inverse
temperature  $\beta=1/kT$:
 \begin{equation}
\sigma\{p_{pmn},u_{pmn}\}=Z^{-1}\exp\bigg[-\beta H\{p_{pmn},u_{pmn}\}\bigg]
\label{16}
 \end{equation}
where $Z$ is the normalizing constant. Now, if we express, using Eq. (\ref{9}), the changes of the cavity
thickness $\xi(t,x,y)$ through the degrees of freedom $u_{pmn}$
\begin{equation}
\xi(t,x,y)=u(t,x,y,0)-u(t,x,y,L)=2\sum_{pmn} u_{pm,2n-1}\cos\bigg({\pi p x\over a}\bigg)\hskip1mm
\cos\bigg({\pi m y\over b}\bigg),\hskip3mm p,m,n>0,
\label{17}
\end{equation}
then, for the correlation function entering Eq. (\ref{5}), we obtain the expression
\begin{equation}
\langle\xi(0,x,y)\xi(t,x',y')\rangle=
\label{18}
\end{equation}
$$
=4\sum_{pmn} \sum_{p'm'n'}\langle u_{pm,2n-1}(0)
u_{p'm',2n'-1}(t)\rangle
\cos\bigg({\pi p x\over a}\bigg)\hskip1mm
\cos\bigg({\pi m y\over b}\bigg)
\cos\bigg({\pi p' x'\over a}\bigg)\hskip1mm
\cos\bigg({\pi m' y'\over b}\bigg)
$$
Since the distribution function is factorized over the degrees of freedom $u_{pnm}$
 (i.e., the degrees of freedom
are independent random quantities with zero mean value), the off-diagonal mean values in the double sum
of Eq. (\ref{18}) vanish, and we have
\begin{equation}
\langle\xi(0,x,y)\xi(t,x',y')\rangle=
\label{19}
\end{equation}
$$
=4\sum_{pmn} \langle u_{pm,2n-1}(0)
u_{pm,2n-1}(t)\rangle
\cos\bigg({\pi p x\over a}\bigg)\hskip1mm
\cos\bigg({\pi m y\over b}\bigg)
\cos\bigg({\pi p x'\over a}\bigg)\hskip1mm
\cos\bigg({\pi m y'\over b}\bigg)
$$

Let us calculate the correlator of the type $\langle u(0)u(t)\rangle$
 (the subscripts are omitted for brevity). Since the
degrees of freedom meet the equation of motion $\ddot u=-\omega^2 u$,
 we can write the following expression for $u(t)$:
\begin{equation}
u(t)=u(0)\cos[\omega t]+{\dot u(0)\over \omega}\sin[\omega t]\Rightarrow
\langle u(0)u(t)\rangle=\langle u^2(0)\rangle\cos[\omega t]
+\langle u(0)\dot u(0)\rangle{\sin[\omega t]\over \omega}
\label{20}
\end{equation}
The velocity $\dot u(0)$ entering the last term of Eq. (\ref{20})
 is proportional to the corresponding generalized
momentum, which is a random quantity independent of $u$. Therefore, $\langle u(0)\dot u(0)\rangle=0$,
 and we have:
\begin{equation}
\langle u(0)u(t)\rangle=\langle u^2(0)\rangle\cos[\omega t]
\label{21}
\end{equation}
It follows from Eqs. (\ref{15}) and (\ref{16}) that, if we introduce notation $\alpha\equiv M\omega^2/16 kT$,
 then
\begin{equation}
\langle u^2(0)\rangle=\sqrt{\alpha\over\pi}\int u^2\exp [-\alpha u^2]du=
-\sqrt{\alpha\over\pi}{\partial\over\partial\alpha}\int \exp [-\alpha u^2]du={1\over 2\alpha}=
{8kT\over M\omega^2}
\label{22}
\end{equation}
Thus, using Eq.(\ref{19}), we  obtain
\begin{equation}
\langle\xi(0,x,y)\xi(t,x',y')\rangle=
{32 kT\over M}\sum_{pmn}{\cos[\omega_{pm,2n-1}t]\over \omega_{pm,2n-1}^2}\cos\bigg({\pi p x\over a}\bigg)\hskip1mm
\cos\bigg({\pi m y\over b}\bigg)
\cos\bigg({\pi p x'\over a}\bigg)\hskip1mm
\cos\bigg({\pi m y'\over b}\bigg)
\label{23}
\end{equation}
Now, let us take into account that the dimensions $a$ and $b$ are supposed to be large, i.e., $a,b>>L$.
In this case, we can pass from summation over $p$ and $m$ to integration. By denoting
$[\pi p/a]\equiv A$ and $[\pi m/b]\equiv B$, we obtain
 $$
\omega_{pmn}\rightarrow \omega_n(A,B)=v\sqrt{A^2+B^2+\bigg({\pi n\over L}\bigg)^2},\hskip3mm
 dA=[\pi/a],\hskip3mm
 dB=[\pi/b]
$$
and
\begin{equation}
\langle\xi(0,x,y)\xi(t,x',y')\rangle=
{32 kT\over M}{ab\over\pi^2}\sum_{n}\int_0^\infty dAdB
{\cos[\omega_{2n-1}(A,B)t]\over \omega_{2n-1}^2(A,B)}\cos(Ax)\hskip1mm
\cos(By)\cos(Ax')\hskip1mm\cos(By')
\label{24}
\end{equation}
Since Eq. (\ref{5}) includes the correlation function (\ref{24})
averaged over the area of the light spot, it is
convenient to introduce the function $F(A,B)$ defined as
 \begin{equation}
 F(A,B)\equiv\int_{D^2} dxdy\cos[Ax]\cos[By]
 \label{25}
 \end{equation}
Then, taking into account that
  $$
  \int\cos[\Omega t]e^{\imath\nu t}dt=\pi[\delta(\nu-\Omega)+\delta(\nu+\Omega)],
  $$
and using Eqs.  (\ref{4}), (\ref{24}), and (\ref{25}), we obtain for the
sought spectral density of the noise $S(\nu)$ the following equation

  \begin{equation}
  S(\nu)=G^2(\omega){32 kT\over M}{ab\over\pi}\sum_{n}^{odd}\int_0^\infty dAdB
{\delta[\nu-\omega_{n}(A,B)]\over \nu^2} F^2(A,B)
  \label{27}
  \end{equation}
Here, the symbol {\it odd} shows that summation is performed over odd $n$.
Using the presence of $\delta$-function, we can perform integration over $B$, and eventually we have
  \begin{equation}
  S(\nu)={G^2(\omega)\over \nu}{32 kT\over L\rho\pi v^2}\sum_{n}^{odd}\int dA
\bigg[{\bigg({\nu\over v}\bigg)^2-A^2-\bigg({\pi n\over L}\bigg)^2}\bigg]^{-1/2}
F^2\bigg(A,\sqrt{\bigg({\nu\over v}\bigg)^2-A^2-\bigg({\pi n\over L}\bigg)^2}\bigg)
  \label{28}
  \end{equation}

Here $\rho\equiv M/abL$ is the density of the cavity material,
 and the integration is performed over the range of
values of the variable $A$, where the radical entering Eq. (\ref{28}) is real.
 The contribution of the $n$-th
mode is evidently nonzero only at $\nu>[\pi n v/L]$.
 As was already mentioned, the greatest value of $G(\omega)$ is achieved
when the interferometer is tuned to the slope of the resonance:
$$
G_{max}=G(\omega_0+\Delta)={I_0Q\over 2LD^2}.
$$
In this case, the noise intensity is the greatest and equals
 \begin{equation}
  S(\nu)=
  {(I_0Q)^2\over \nu L^3D^4}\hskip1mm{8 kT\over \rho\pi v^2}\sum_{n}^{odd}\int dA
\bigg[{\bigg({\nu\over v}\bigg)^2-A^2-\bigg({\pi n\over L}\bigg)^2}\bigg]^{-1/2}
F^2\bigg(A,\sqrt{\bigg({\nu\over v}\bigg)^2-A^2-\bigg({\pi n\over L}\bigg)^2}\bigg)
  \label{280}
  \end{equation}
 To make quantitative estimates, we  assume, for simplicity, that the light beam has a square cross section, i.e.,
$x\in[a/2-D/2,a/2+D/2]$ and $y\in[b/2-D/2,b/2+D/2]$.
 In this case, the function $F(A,B)$ (\ref{25}) can be obtained in the explicit form:
$$
F(A,B)={4\over AB}\cos\bigg({Aa\over 2}\bigg)\sin\bigg({AD\over 2}\bigg)
\cos\bigg({Bb\over 2}\bigg)\sin\bigg({BD\over 2}\bigg)
$$

Note that the rapidly oscillating terms (of the type $\cos^2[Bb/2]$)
 arising in the formula for $F^2(A,B)$ can be replaced
by their mean values (i.e., by 1/2). With allowance for these remarks, we may accept that
 \begin{equation}
 F^2(A,B)={1\over A^2B^2}\sin^2\bigg({AD\over 2}\bigg)
\sin^2\bigg({BD\over 2}\bigg)
\label{29}
 \end{equation}

All the above calculations refer to infinite phonon lifetime in the cavity material.
This is the reason why the correlation function (\ref{21}) does not decay. If we take into account the decay
of the acoustic mode free vibrations and temporal evenness of the correlation function, we obtain, for
the correlator, the following expression:
\begin{equation}
\langle u(0)u(t)\rangle=\langle u^2(0)\rangle\cos[\omega t]\exp-\bigg|{t\over\tau}\bigg|
\label{30}
\end{equation}
where $\tau$ -- is the phonon lifetime. The noise spectrum $S_\tau(\nu)$,
 in this case, is obtained as a convolution of
Eq.(\ref{280}) and Lorentzian with the width
 \footnote{As we will
see below, the noise spectrum appears to be localized in a rather narrow spectral region, where the
frequency dependence of $\tau$ may be neglected}
equal to the inverse phonon lifetime  $\tau^{-1}$:
 \begin{equation}
S_\tau(\nu)={\tau\over \pi}\int {S(\nu'-\nu)\over 1+[\nu'\tau]^2}\hskip1mmd \nu'
 \label{31}
 \end{equation}

 \section{Possibility of observation of the microcavity noise}

One of the most popular optical microcavities is the Bragg cavity, namely, the Fabry-Perot
interferometer comprised of two Bragg mirrors separated by a half-wave gap. In spite of the fact that
these microcavities represent multi-layer structures, the intensity noise of the transmitted light can
be estimated using Eqs. (\ref{280}) and (\ref{31}) for the following reason.
It is essential for the single-layer
model considered above that both the optical and the acoustic modes are localized in the {\it same layer} of
the material which may serve as a resonator both for the optical and for the acoustic waves. A similar
situation may take place in real Bragg cavities, since the optical Bragg mirror also has an acoustic
stop-band and can efficiently reflect acoustic waves with the appropriate frequencies. In this case,
the half-wave gap (for optical waves) between the Bragg mirrors can form an acoustic resonator, whose
properties can be approximately described by the single-layer model described in the previous section.
Our quantitative estimates show that, for a typical half-wave Bragg cavity with
$\lambda_0=2\pi c/\omega_0=800$ nm, comprised
of the $TiO_2$ and $SiO_2$ layers, the frequency of the lowest acoustic mode equals 10 GHz and matches the
acoustic stop-band of the Bragg mirrors. The possibility of manufacturing of efficient acoustic Bragg
resonators was demonstrated in \cite{Lan,Pas,Lan1,Trigo}.

Bearing in mind all the aforesaid, let us accept the following values
 of the parameters entering Eqs. (\ref{280})
and (\ref{31}): $L=0.276 \hskip1mm\mu$m, $I_0=0.1$ W, $Q=1000$, $v=5570$ m/s ($SiO_2$), $\rho=2000$ kg/m$^3$
($SiO_2$).
For the above values of the parameters, the frequency $\nu_1$ of the lowest acoustic mode ($n
=1$) can be estimated as $\nu_1=v/2L\approx 10$ GHz. As seen from Eqs. (\ref{280}) and (\ref{31}),
  the noise power increases with
decreasing size of the light spot $D$. In our calculation, we accept that $D = 10 \hskip1mm\mu$m.
 This size of the
spot does not contradict the above finesse $Q=1000$.
\footnote{Futher decrease of $D$ by light focusing can
be accompanied by a decrease of the finesse, which occurs due to increasing uncertainty in the angle of
incidence. However, for normal incidence of the focused beam, this effect is not strong. }
To estimate the decay time $\tau$
 of acoustic vibrations, entering Eq. (\ref{31}), we can take into account that the
finesse of the quartz resonators at frequencies of around $10^8$ Hz may reach $10^4 - 10^5$. For the
frequencies $\sim 10^{10}$ Hz, we are interested in, the finesse of the acoustic vibrations is expected to be
lower. Thus, for our estimates, we will accept the finesse of the acoustic mode to be $\sim$2000. In this
case, the quantity $\tau$ is determined by the relationship $\nu_1\tau=2000$.
 The possibility of manufacturing
of acoustic microcavities with a finesse of around 1000 in the THz range was reported in \cite{Rozas}.

To estimate possibility of observation of the intensity noise of the light transmitted by the
microcavity, the magnitude of this noise (determined by Eqs (\ref{280}) and (\ref{31}))
  should be compared with the
shot noise of the light source
 \begin{equation}
 S_{sn}=I_0\hbar\omega_0
 \label{32}
 \end{equation}

The results of this comparison are shown in Fig. 1. This figure shows the calculated noise spectra in
the region of the lowest acoustic mode ($n=1$) for the infinite (oscillating curve) and finite (smooth
curve) phonon lifetime. Horizontal line shows the level of shot noise of the used light Eq. (\ref{32}).
Since observation of the light shot noise with the aid of the up-to-date spectrum analyzers does not
encounter any problems, then, as seen from Fig. 1, we have all the grounds to believe that the above effect
can be detected even if the results of our calculations appear to be overestimated by 1-2 orders of
magnitude. \footnote{Spectral region of the modern digital spectrum analyzers is restricted to the
frequencies of around 1-2 GHz. For this reason, for detection of the noise signal with the frequencies
~ 10 GHz, considered in this paper, one has to perform the appropriate transfer of the spectrum. A
similar task is solved in the systems of satellite TV with the aid of heterodyne converters, which can
also be used in this case. }

\section{Conclusions}

In this paper, we presented calculations of spectral power density of the light transmitted by a
microcavity. It is shown that this power density reveals a peak at the frequency of acoustic vibrations
of the cavity -- the effect similar to Raman scattering. The quantitative estimates are made which show
that the noise arising due to the mechanism considered in the paper can be detected using the
up-to-date noise-spectroscopy technique.

Without entering into details of possible informative capabilities of this technic, note only that
 thermal
vibrations are usually considered as a spurious factor that restricts operational
 stability of devices (see, i.g.,\cite{Javan}. The above calculations show, however, that
 the intensity noise spectrum of the light transmitted by
the cavity contains information about the spectrum of acoustic vibrations of the structure. Measuring
variation of this spectrum versus the light spot diameter will allow one to judge about plausibility of the
used simple model which neglects, in particular, disorder of the real structure and possible localization of the
acoustic waves. Observation of the correlation function of the noise for two spaced light beams will
possibly allow one to evaluate the localization radius of acoustic vibrations of the structure.

\section{Acknowledgements}

The author is grateful to V.G.Davydov for useful discussions.

\
\newpage
\begin{figure}
\begin{center}
\includegraphics [width=10cm]{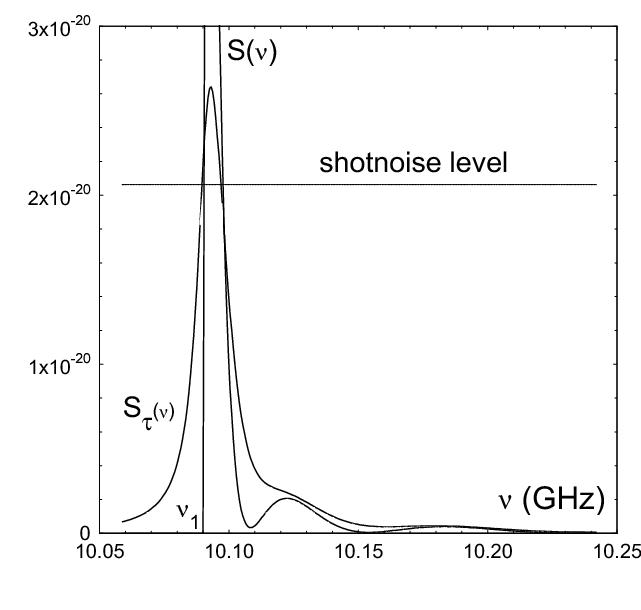}
 \caption{The intensity noise spectrum of the light transmitted by the cavity. The values of the parameters are:
thickness of the microcavity $L = 0.276\hskip1mm\mu$m, finesse of
the transmission spectrum $Q = 1000$, and the light beam intensity
$I_0 = 0.1$ W. The calculation is performed for the spectral
region of the lowest acoustic mode of the cavity $\nu_0\approx 10$
GHz. Oscillating curve $S(\nu)$ corresponds to the noise spectrum
at infinite phonon lifetime. Smooth curve  $S_\tau(\nu)$ -- the
same at $2\pi\nu_0\tau=2000$.
 Horizontal line shows the shot noise level of the light.}
 \label{fig1}
 \end{center}
\end{figure}

\end{document}